%% file: main.tex
\documentclass[10pt,conference,anonymous]{IEEEtran}
\usepackage{graphicx} % Required for inserting images
\usepackage{amsmath,amssymb,amsfonts}
\usepackage{textcomp}
\usepackage{xcolor}
\usepackage{stmaryrd}
\usepackage{algorithm}
\usepackage{algpseudocode}
\usepackage{mathtools}
\usepackage{parcolumns}
\usepackage{listings}
\usepackage{xcolor}
\usepackage{booktabs}
\usepackage{parcolumns}
\usepackage{xspace}
\usepackage{url}
\usepackage[T1]{fontenc}

\input{macros.tex}

\colorlet{punct}{red!60!black}
\definecolor{background}{HTML}{EEEEEE}
\definecolor{delim}{RGB}{20,105,176}
\colorlet{numb}{magenta!60!black}
\lstdefinelanguage{json}{
    basicstyle=\scriptsize\ttfamily,
    showlines=true,
    frame=tlrb,
    backgroundcolor=\color{background},
}

\begin{document}

\title{Synthesizing Access Control Policies using Large Language Models}

\author{
\IEEEauthorblockN{1\textsuperscript{st} Adarsh Vatsa}
\IEEEauthorblockA{\textit{Department of Computer Science} \\
\textit{Stevens Institute of Technology}\\
Hoboken, USA \\
avatsa@stevens.edu}
\and
\IEEEauthorblockN{2\textsuperscript{nd} Pratyush Patel}
\IEEEauthorblockA{\textit{Department of Computer Science} \\
\textit{Stevens Institute of Technology}\\
Hoboken, USA \\
ppatel10@stevens.edu}
\and
\IEEEauthorblockN{3\textsuperscript{rd} William Eiers}
\IEEEauthorblockA{\textit{Department of Computer Science} \\
\textit{Stevens Institute of Technology}\\
Hoboken, USA \\
weiers@stevens.edu}
}

\maketitle

\begin{abstract}
Cloud compute systems allow administrators to write access control policies that govern access to private data. While policies are written in convenient languages, such as AWS Identity and Access Management Policy Language, manually written policies often become complex and error prone. In this paper, we investigate whether and how well Large Language Models (LLMs) can be used to synthesize access control policies. Our investigation focuses on the task of taking an access control request specification and zero-shot prompting LLMs to synthesize a well-formed access control policy which correctly adheres to the request specification. We consider two scenarios, one which the request specification is given as a concrete list of requests to be allowed or denied, and another in which a natural language description is used to specify sets of requests to be allowed or denied. We then argue that for zero-shot prompting, more precise and structured prompts using a syntax based approach are necessary and experimentally show preliminary results validating our approach. 
\end{abstract}


\section{Introduction}
Modern cloud computing platforms allow customers to secure their data through \textit{access control policies}. Most cloud computing services have their own access control specification language for writing policies.
The goal of crafting an access control policy is to create rules adhering to specifications for allowing and denying access while minimizing requests, or permissions, required to complete a given task. Manually crafting access control policies is tedious and often error prone, as ensuring the policy gives enough access to perform a given task while simultaneously controlling the correct level of permissions is difficult and requires immense domain-specific knowledge. 

%Moreover, modifying existing policies to accommodate new access control requests, or repairing a policy to remove access control requests, is error-prone and can result in unintentionally allowing access to secure or confidential data. Thus, developing approaches for not just analyzing access control policies, but also for crafting access control policies, must be verification-driven.
Large Language Models (LLMs) have shown exceptional capabilities in code
generation and summarization, highlighting the lucrative prospects particularly in automation
and software engineering. 
%More recently, LLMs have been used in software engineering such as automated program repair and automated test generation. 
% However, despite the impressive capabilities of LLMs they are seen as largely unreliable, often
% hallucinating or generating incorrect results. Moreover, LLMs lack explainability in their results. Even when LLMs generate correct results, it is currently impossible to determine why or how the LLM gave the result that it did. 
% Thus when using LLMs, it is imperative to have techniques which can analyze the correctness of their results.
In this work, we investigate how to leverage large language models for policy synthesis in a verification-based approach. Given a request specification as input, we leverage the natural language capabilities of large language models to automatically synthesize an access control policy which correctly adheres to the request specification. Our work aims to answer the following research questions: \\
\textbf{RQ1: Can LLMs synthesize valid access control policies from a given whitelist and blacklist set of requests?} \\
\textbf{RQ2: Can LLMs correctly synthesize access control policies given a natural language description of the allowed or denied requests?} \\
\textbf{RQ3: Can well-crafted, more structured prompts improve the capability of LLMs in generating correct policies?} \\
The rest of the paper is structured as follows. In Section~\ref{sec:related} we discuss the work most related to our approach. In Section~\ref{sec:analyzing_policies} we briefly overview access control policies and an approach to verify them. In Section~\ref{sec:method} we discuss the dataset we use in our approach, and discuss our process for synthesizing and evaluating access control policies using large language models. In Section~\ref{sec:experiments} we discuss our experimental evaluation. Finally, in Section~\ref{sec:conclusion} we conclude the paper.
 
\section{Related Work}\label{sec:related}
% Access control~\cite{samarati01policiesmodelsmechanisms,sandhu94access,sandhu96audit}, access control policy languages~\cite{abadpeiro99plas,jajodialogical,jajodia01multiplepolicies,jajoida97unified}, and verification approaches~\cite{DBLP:conf/cade/DoughertyFK06,
% DBLP:journals/sttt/HughesB08,schaad:lightweight,zao:rbac} has been the subject of extensive research. 
There is existing work for assisting the creation of policies~\cite{DBLP:conf/sacmat/FislerK10,
Egelman:2011:OID:1978942.1979280} but these techniques lack automated techniques for policy creation with verifiable guarantees. Policy mining is another popular approach~\cite{mining1,mining2} and has been used to automatically generate policy specifications from a given set of allowed or denied accesses. In~\cite{subramaniam2024intentbasedaccesscontrolusing,enforcingaccessusingllms} the authors propose  LLMs as a coding assistant to help administrators write software enforcing access control. Constraint solving approaches for policy analysis has been a subject of recent research, particularly in the proprietary Zelkova tool used within AWS~\cite{BBC18} which translates policies into SMT formulas. The work in~\cite{ESL22,Quacky23} extended this by introducing quantitative analysis for reasoning about the permissiveness of policies, and was recently applied to quantitative policy repair~\cite{QuackyRepair}. 

\section{Analyzing Access Control Policies}\label{sec:analyzing_policies}
An access control policy specifies \textit{who} can do \textit{what} under \textit{which} conditions. Access control policies consist of sets of rules specifying which \textit{principals} can perform which \textit{actions} on which \textit{resources} under which \textit{conditions}. Given an access request and associated policy, access is granted \textit{if and only if} there exists a rule in the policy allowing access and no rules in the policy explicitly deny access.

Given an access control policy $\P$, one can reason over the requests allowed by $\P$ by encoding the set of requests allowed by $\P$ into a Satisfiability Modulo Theories formula $\smtP$ where satisfying solutions to $\smtP$ correspond to requests allowed by $\P$. For full details of this encoding, we refer the reader to~\cite{ESL22,Quacky23}. The permissiveness of $\P$ is a measure of how many requests are allowed by $\P$. Thus, to compute the permissiveness of $\P$, one must count the number of requests allowed by $\P$, which corresponds to computing $\mcount{\smtP}$, the number of satisfying solutions to $\smtP$. Given two policies $\Pone$, $\Ptwo$, we can compare the requests allowed by checking the satisfiability of $\smtPone \land \neg \smtPtwo$ (requests allowed by $\Pone$ but not $\Ptwo$) and $\neg \smtPone \land \smtPtwo$ (requests allowed by $\Ptwo$ but not $\Pone$):
\begin{itemize}
    \item If both $\smtPone \land \neg \smtPtwo$ and $\neg \smtPone \land \smtPtwo$ are unsatisfiable, then they allow the same set of requests
    \item If $\smtPone \land \neg \smtPtwo$ is satisfiable but $\neg \smtPone \land \smtPtwo$ is unsatisfiable, then $\Pone$ accepts all the requests $\Ptwo$ does and more, so we say that $\Pone$ is strictly more permissive than $\Ptwo$
    \item If $\smtPone \land \neg \smtPtwo$ is unsatisfiable but $\neg \smtPone \land \smtPtwo$ is satisfiable, then $\Ptwo$ accepts all the requests $\Pone$ does and more, we say that $\Ptwo$ is strictly more permissive than $\Pone$
    \item If both are satisfiable, then both $\Pone$ and $\Ptwo$ accept different sets of requests and so we say they are incomparable.
\end{itemize}
By computing $\mcount{\smtPone \land \neg \smtPtwo}$ (the number of requests allowed by $\Pone$ but not by $\Ptwo$) and $\mcount{\neg \smtPone \land \smtPtwo}$ (the number of requests allowed by $\Ptwo$ but not by $\Pone$) one can quantify the relative permissiveness between $\Pone$ and $\Ptwo$. Computing the count can be done by a model counting constraint solver.

\section{Methodology}\label{sec:method}
In this section we outline our policy datasets, the access control request specifications, how we evaluate the synthesized policies, and our experiment configuration.

\subsection{Policy Dataset}
To the best of our knowledge, there is no existing dataset for policy analysis 
nor prompting methodology for their creation using large language models. Moreover, there is a shortage of existing policy datasets for research purposes~\cite{Quacky23} and recent research on access control policies has been done on closed-source datasets~\cite{BBC18,DAntoni2024}.
% In this work, we extend the policies from the Quacky dataset~\cite{ESL22,Quacky23} with several more policies, and created prompts for them based on realistic access control scenarios.
% To query the LLM to synthesize an access control policy, we constructed prompts where each prompt gives the LLM an access request specification and ask the LLM to create a policy which adheres to the specification, where the policy is the set of access control rules implemented in a specific access control policy language. For this work, we focus on policies written in the AWS IAM language (although our approach is not specific to AWS policies). 
Thus, to investigate how well LLMs can generate access control policies, we consider the initial set of 40 policies from Quacky~\cite{Quacky23} dataset and extend the dataset with 7 more synthetically generated policies. The 40 policies from Quacky were taken from AWS user forums and consisted of policies that users posted to receive feedback or troubleshoot their policies. We extended this dataset by manually crafting 7 more policies which were similar in complexity but varied in their intention: from policies allowing or denying certain roles and principals in accessing objects within specific S3 buckets, to policies managing EC2 instances or the creation and/or deletion of IAM roles. This set of 47 policies consists of the \textit{ground truth} policies from which we will use to compare with the synthesized policies created from the LLM.

\subsection{Request Specifications}
To query the LLM to synthesize an access control policy, we constructed prompts where each prompt gives the LLM an access request specification and ask the LLM to create a policy which adheres to the specification, where the policy is the set of access control rules implemented in a specific access control policy language. For this work, we focus on policies written in the AWS IAM language (although our approach is not specific to AWS policies). The prompts we used to query the LLM to generate policies each consist of an access request specification which details sets of requests which should be allowed or denied.
We consider three types of prompts, each varying in the specificity with how requests should be allowed or denied, corresponding to each of our three research questions. As existing prompts for policy generation do not exist, we detail how we constructed each type of prompt.

\begin{figure}[t]
\begin{lstlisting}[language=json, frame=tlrb]
Create an AWS IAM policy that incorporates all of the 
following requests. Return only the JSON policy, 
nothing else:
...
{"principal": "alice",
 "action": "s3:PutObjectAcl",
 "resource": "mybucket/backups/data/file8.txt"
}
...
\end{lstlisting}
\vspace{-1.8mm}
\begin{lstlisting}[language=json, frame=tlrb]
Create an AWS IAM policy based on this description. 
Return only the JSON policy, nothing else:
...
Requests by Alice to read objects in the public 
bucket should be allowed.
...
\end{lstlisting}
\vspace{-1.8mm}
\begin{lstlisting}[language=json, frame=tlrb]
Create an AWS IAM policy based on this description. 
Return only the JSON policy, nothing else:
...
ALLOW user:alice READ bucket:public-bucket/
*Note: Use ACCOUNT_ID as placeholder in ARNs
...
\end{lstlisting}
\caption{Example of a concrete-request prompt (topmost, (a)), a natural-language prompt (middle, (b)), and a fine-grained with syntax prompt (bottom, (c)), for the same policy. Only snippets of the prompts are shown.}
\label{fig:prompt_example}
\end{figure}

\subsubsection{Concrete-Request}
These prompts consist of specific requests which should be allowed or denied access. Figure~\ref{fig:prompt_example}(a) shows an example of a concrete-request prompt. The request generation process creates synthetic AWS S3 access patterns by randomly combining predefined S3 actions (such as GetObject, PutObject, etc.) with dynamically generated resource paths. Each generated dataset contains between 30-150 allowed requests and 5-20 denied requests, where each request specifies a principal, an action, and a resource path. The resource paths are constructed by combining random selections from predefined buckets, directories, and file names, with varying depths to simulate realistic S3 storage hierarchies.

\subsubsection{Coarse-Grained}
These prompts use a less-structured natural language description giving a general idea of what requests should be allowed or denied. Unlike concrete-request prompts, coarse-grained prompts give a loose description detailing how sets of requests should be allowed or denied. Figure~\ref{fig:prompt_example}(b) shows an example of a coarse-grain prompt. We synthesized 47 prompts, each for one ground-truth policy, by taking each policy and having a software engineer write a general description of requests which are allowed or denied by the ground-truth policy.
    
\subsubsection{Fine-Grained-with-Syntax}
%These prompts combine natural language description with SQL inspired syntax using structured, precise language for how requests should be allowed or denied access. Unlike the coarse-grain prompts, these prompts use specific language  to guide the LLM in synthesizing policies better capturing access request specification. Figure~\ref{fig:prompt_example}(c) shows an example of one such prompt. We synthesized 47 prompts, one for each ground-truth policy, by taking each policy and having a software engineer write a detailed, precise description of how sets of requests should be allowed or denied.
These prompts implement a SQL-inspired structured syntax for specifying access control requirements. This approach moves beyond natural language toward a more formal policy specification language, providing clear, unambiguous instructions to guide the LLM in synthesizing accurate IAM policies. We synthesized 47 prompts, each corresponding to a ground-truth policy, where a software engineer crafted specifications using this structured format. 

\subsection{Evaluating Synthesized Policies}
We consider two scenarios in which a cloud administrator queries an LLM to generate an access control policy. In the first scenario, the LLM is queried to generate a policy from a combined whitelist/blacklist set of requests (Concrete-Request prompts). This scenario will determine how well an LLM can synthesize a policy from request specifications consisting of a concrete list of requests which the policy should allow or deny access. In the second scenario, the LLM is queried to generate a policy from a natural language description of the allowed and denied requests.
We now discuss our approach to verify the correctness of the policy synthesized for each case.

\paragraph{Verifying Concrete-Request Synthesized Policies} In this scenario, the LLM is given a list of specific access requests which should be allowed or denied, and is queried to generate a policy from them. To verify the correctness of the synthesized policy, we encoded each request into an SMT formula and checked it against the policy's encoded formula, recording whether access was correctly allowed or denied (as in ~\cite{ESL22}). We repeat this process for every access request. While this method is slow, it produces the same results as if we used an AWS policy simulator to check each request against the policy. The correctness of the policy synthesized by the LLM for one prompt is then $\frac{\text{\# correctly classified requests}}{\text{\# total requests}}$.

\paragraph{Verifying Coarse-Grained and Fine-Grained-with-Syntax Synthesized Policies} In this scenario, the prompts consist of natural language description of the allowed or denied requests. We conducted two experiments for this scenario, one in which we prompted the LLM using the coarse-grained request specification prompts, and one in which we prompted the LLM using the more precise fine-grained-with-syntax request specification prompts. We queried the LLM with each prompt and had it synthesize an access control policy which adheres to the access request specification within the prompt. To verify the correctness of the synthesized policy, we compare the synthesized policy from the LLM with the ground-truth policy by comparing the relative permissiveness of both policies (see Section~\ref{sec:analyzing_policies}). \textbf{If the synthesized policy and the original policy are equivalent, then the LLM is correct for that instance}. Otherwise, we determine if either is more permissive than the other, or incomparable.

\section{Results}\label{sec:experiments}
In this section we discuss our experimental results and how they answer our three research questions. We used GPT-4 as our large language model, and all of our prompts are zero-shot. We ran all of our experiments~\footnote{available at \url{https://github.com/Saffronius/Synthesizing_ACP}} on a Windows 11 computer with a GeForce RTX 4090 GPU with 16GB Vram, Intel Core i-14900HX 2.20Ghz, and 32GB RAM. To translate policies into constraint formula and compare policies, we used the latest version of the Quacky tool~\cite{Quacky23}.

\begin{figure}[t]
    \centering
    \includegraphics[width=\linewidth]{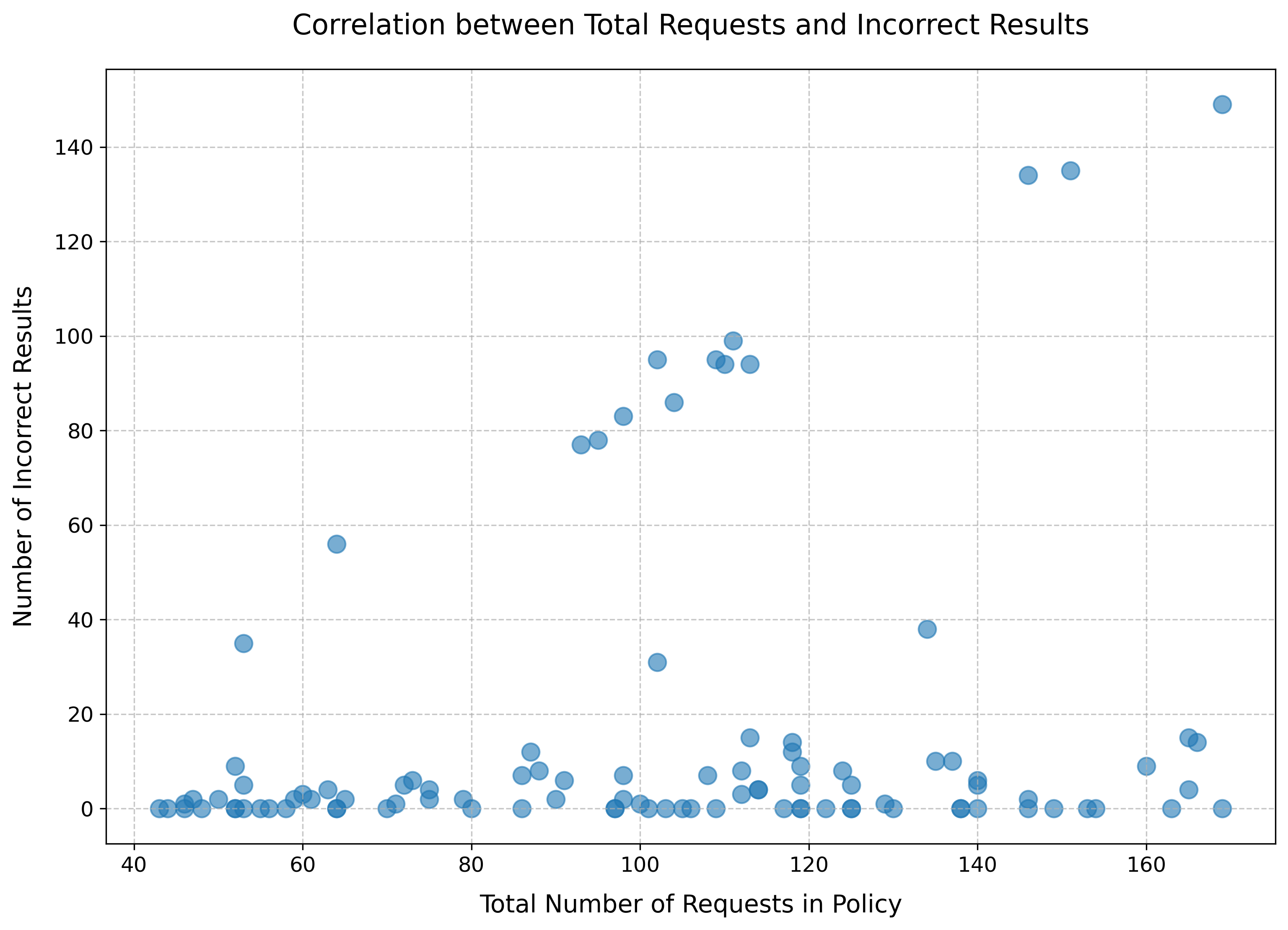}
    \caption{Number of incorrectly classified requests compared to the number of total requests for each synthesized policy for concrete-request prompts. }
    \label{fig:exp1}
\end{figure}

\subsection{RQ1: Synthesizing Policies from Concrete-Request Specifications} We queried the LLM to synthesize 100 policies for 100 request specifications. We varied the number of requests for each prompt from 35 requests to 180 requests and recorded how many of the requests were correctly classified by the synthesized policy in each case. The results are shown in Figure~\ref{fig:exp1}. All the synthesized policies were syntactically correct with an average classification rate of $83.82\%$, though the LLM occasionally had difficulty generating correct policies. Thus, we found that in general the LLM did well in synthesizing policies from lists of requests, but can experience difficulty when the number of request is large and there is a large variety in the types of requests to be allowed or denied.

\subsection{RQ2: Synthesizing Policies from Coarse-Grain Request Specifications} 
\begin{figure}[t]
    \centering
    \includegraphics[width=\linewidth]{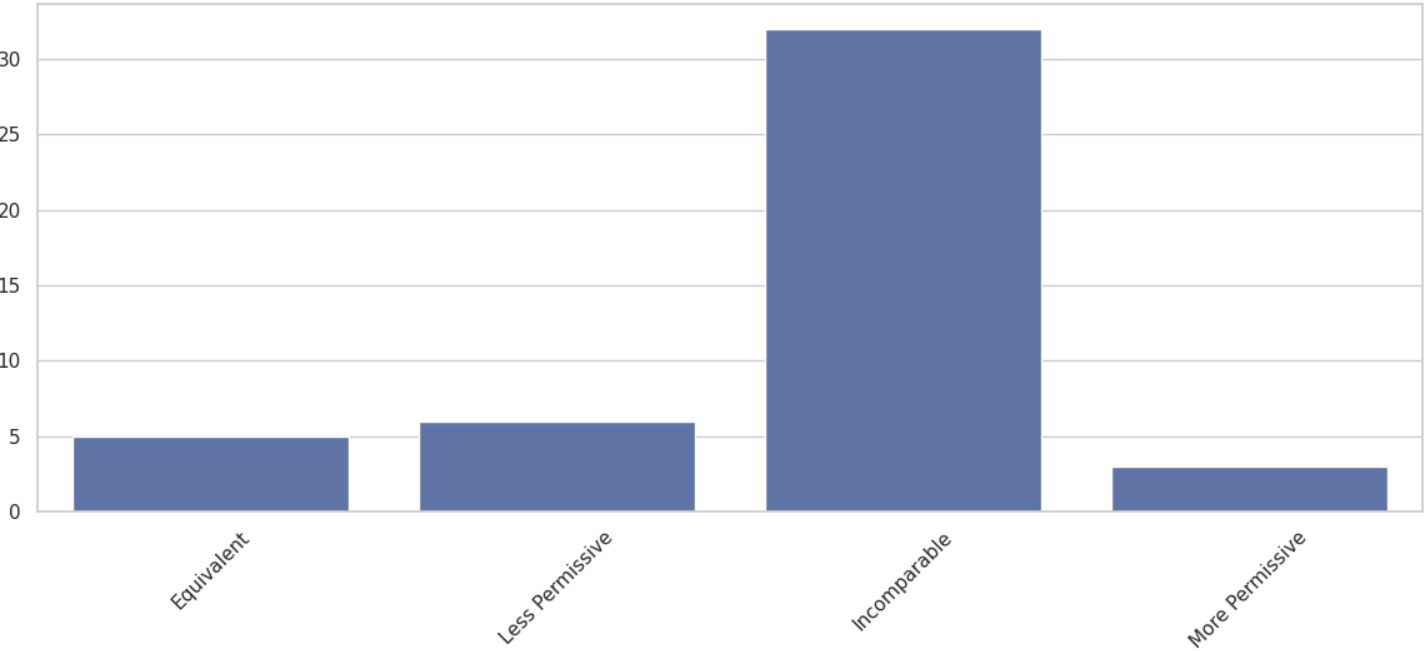}
    \caption{Distribution of how the synthesized policy compare to the ground truth policy for coarse-grain request prompts. }
    \label{fig:exp2}
\end{figure}

We queried the LLM to synthesize 47 policies for 47 requests specifications where each request specification consisted of a loose natural language description of the requests to be allowed or denied. We compared the sets of requests allowed by each synthesized policy to the ground-truth policy and reported on the relative permissiveness between them. The results are shown in Figure~\ref{fig:exp2}. In most cases the LLM synthesized a policy that was incomparable to the ground-truth policy. Manual inspection showed the synthesized policies were too broad due to imprecise request specifications, causing significant differences from ground-truth policies. Thus, we infer that LLMs can synthesize syntactically valid policies from natural language description of allowed and denied requests, but the resulting policy can differ greatly in permissiveness depending on how the prompt is constructed.

\begin{figure}[t]
    \centering
    \includegraphics[width=\linewidth]{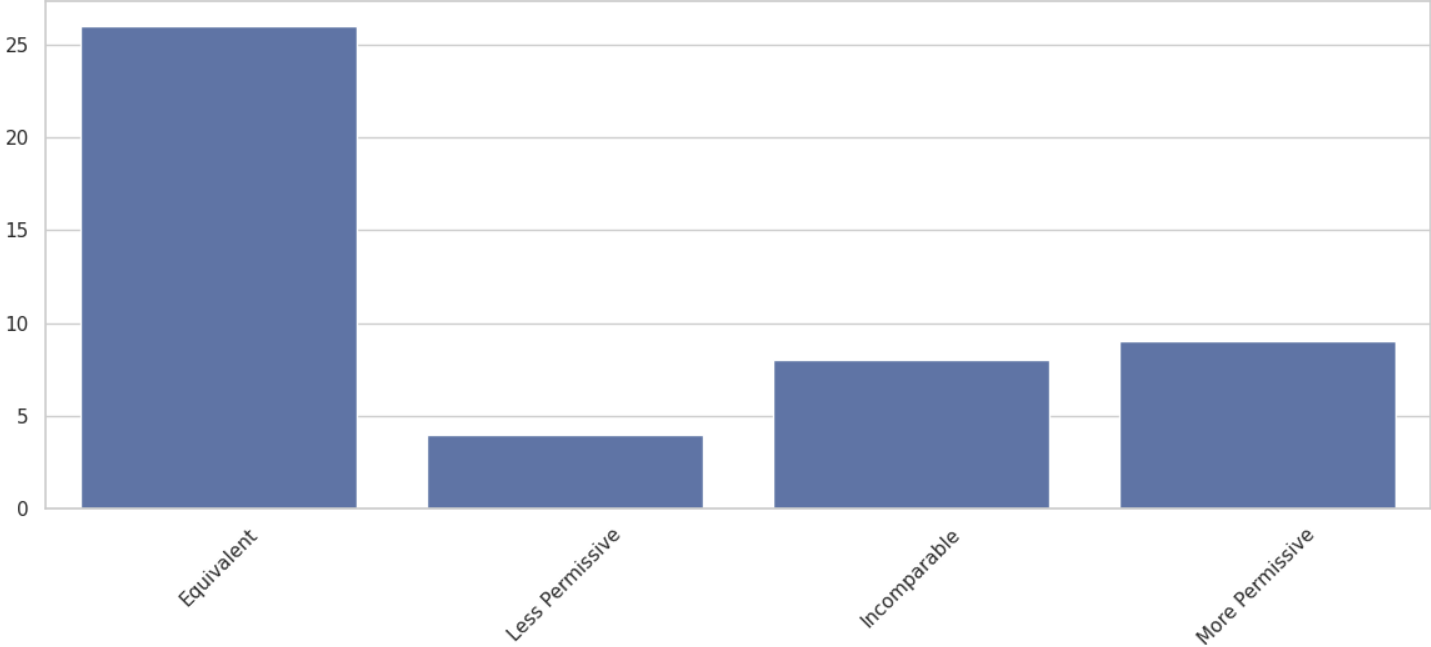}
    \caption{Distribution of how the synthesized policies compare to the ground truth policies for each of the fine-grain-with-syntax prompts.}
    \label{fig:exp3}
\end{figure}

\subsection{RQ3: Synthesizing Policies from Fine-Grain-with-Syntax Request Specifications} We queried the LLM to again synthesize 47 policies for 47 request specifications, this time with more detailed well-structured requests specifications to guide the LLM in generating more correct policies. We again compare the synthesized policies against the ground truth policies and report on the relative permissiveness between them. The results are shown in Figure~\ref{fig:exp3}. We found that by constructing prompts with specific guidance in how requests should be allowed or denied, the LLM was able to give a more precise policy in the majority of cases. \textbf{The high success rate of this approach suggests the potential for developing a dedicated policy specification language that, when combined with a fine-tuned LLM, could serve as a reliable bridge between human-readable access requirements and formally correct IAM policies.} Future work could explore enhancing this syntax with natural language elements while maintaining its precision, potentially creating a hybrid specification language that balances human readability with formal correctness guarantees.

\section{Conclusion}\label{sec:conclusion}
%Writing correct access control policies for a given access request specification is difficult and requires immense manual effort. We investigated how Large Language Models (LLMs) could be used to assist in the creation of access control policies. Our results show that LLMs can efficiently generate syntactically valid access control policies which perform well for simple access request specifications, but require guidance and effort through well-structured prompting in order to generate correct, precise access control policies. In future work we plan on investigating transfer-learning and fine-tuning to consistently generate correct policies based on our prompting language. We are also conducting experiments on much larger datasets of policies with different cloud environments and performing evaluations with different Large Language models.

Writing correct access control policies for a given access request specification requires significant manual effort. We investigated how Large Language Models (LLMs) could assist in the creation of such policies. Our results show that LLMs can efficiently generate syntactically valid policies that perform well for simple specifications but require well-structured prompting to generate precise policies. Future work will investigate transfer-learning and fine-tuning to consistently generate correct policies using our prompting language, while expanding experiments to larger datasets across different cloud environments and language models.
\bibliographystyle{IEEETran}
\bibliography{other,short_bib}

\end{document}

%% file: macros.tex
\def\smtencode#1{\llbracket #1 \rrbracket}
\def\mcount#1{|#1|}

\def\P{\mathbb{P}}

\def\Pone{\mathbb{P}_1}
\def\Ptwo{\mathbb{P}_2}
\def\smtP{\smtencode{\P}}

\def\smtPone{\smtencode{\Pone}}
\def\smtPtwo{\smtencode{\Ptwo}}

\def\actionnamespace#1{\textsc{#1}\xspace}
\def\ec2{\actionnamespace{ec2}}

\def\s3{\actionnamespace{s3}}